\documentclass[prd,twocolumn,preprintnumbers,aps,
reprint,nofootinbib,longbibliography]{revtex4-1}

\usepackage{amsmath,amssymb,amsfonts,bm}
\usepackage{color,graphicx}
\usepackage[caption=false]{subfig}
\usepackage{array}
\usepackage{latexsym}

\definecolor{blue}{rgb}{0, 0.212, 0.616}
\usepackage{hyperref}
\hypersetup{
colorlinks, linktocpage, bookmarksnumbered,
pdfstartview=FitH,
citecolor=blue,
urlcolor=blue,
linkcolor=red,
filecolor=blue
}

\usepackage{cleveref} 

\usepackage[normalem]{ulem}

\newcommand{\nn}{\nonumber}

\def\r{\rho}
\def\rtc{\rho_{tc}}
\def\rtt{\rho_{\tau\tau}}
\def\cL{{\cal L}}
\def\cH{{\cal H}}
\def\cHeff{\cH_\text{eff}}
\def\g{\gamma}
\def\RD{R_{D^{(\ast)}}}
\def\rD{{R_{D}}}
\def\rDs{{R_{D^{\ast}}}}
\def\bctn{b\to c\tau\bar \nu}

\def\bdtn{B \to D^{(\ast)}\tau \bar \nu}

\def\bdln{B \to D^{(\ast)}\ell \bar \nu}

\def\RK{R_{K^{(\ast)}}}

\def\bsll{b\to s\ell^+\ell^-}
\def\bsmm{b\to s\mu^+\mu^-}

\def\bkstmm{B \to K^\ast \mu^+ \mu^-}

\def\bscc{b\to s c\bar c}

\def\Hpm{H^{+}}
\def\mHpm{m_{H^{+}}}

\def\fbi{\text{fb}^{-1}}
\begin{document}

\title{\boldmath Interplay of the charged Higgs effects in $R_{D^{(\ast)}}$, $b\to s \ell^+\ell^-$ and $W$-mass}
\author{Girish Kumar}
\affiliation{
Department of Physics, National Taiwan University, Taipei 10617, Taiwan}


\begin{abstract}
Current data on semileptonic charged- and neutral-current $B$ decays
show  deviations from the predictions of the Standard Model.
It is well known that a charged Higgs boson, belonging to the
two-Higgs doublet model without $Z_2$ symmetry, offers one of the
simplest solution to the charged-current $B$ decays.
We show that this solution naturally induces a negative shift
of $\mathcal{O}(1)$ in the Wilson coefficient ($C_{9\ell}$)
of operator $(\bar s_L\gamma_\mu b_L)(\bar \ell\gamma^\mu \ell)$,
potentially resolving the tension in neutral-current
$B$ decays as well. Interestingly, the lepton universality ratios in
$b\to s \ell^+\ell^-$ decays, in tune with the recent
LHCb result, remain SM-like.
Precision constraints from neutral $B$ and
$K$ meson mixing, decays $B_c\to \tau\bar\nu$, $B\to X_s\gamma$,
and leptonic decays of $\tau$ and $Z$ can be satisfied.
Furthermore, a positive shift in $W$-boson mass, nicely in agreement
with the CDF measurement, is also possible,
requiring the neutral scalars to be heavier
than the charged Higgs but within the sub-TeV region. 
\end{abstract}

\maketitle

\section{Introduction} \label{sec:intro}

It is remarkable that though there already exists irrefutable
experimental evidence
(e.g., baryon asymmetry of the Universe, neutrino masses)
and persuasive theoretical reasons (e.g., naturalness problem,
flavor problem) for
physics beyond the Standard Model (SM), no new physics (NP) particle
has turned up so far at the LHC.
One reason could be that the NP scale is very heavy and beyond LHC reach.
However, in recent years a number of measurements, especially
those associated with semileptonic decays of $B$ mesons,
have been significantly at odds with the SM predictions
and could be telltale sign of sub-TeV scale NP accessible
at the LHC. We discuss one example of such NP---a charged Higgs
boson $(H^+)$ of a few hundred GeV mass, which can help in alleviating
the tension between theory and the current data.

In semileptonic $B$ decays, one set of prominent  anomalies,
persisting for many years and strengthened further by the recent LHCb measurement \cite{LHCb:RD2022},
are in the lepton flavor universality (LFU) ratios
\begin{align}\label{eq:RD_def}
	\RD = \frac{\mathrm{BR}(\bdtn)}{\mathrm{BR}(\bdln)}; ~~\ell=\{e, \,\mu\}.
\end{align}
The current world average by HFLAV \cite{HFLAV:2019otj}, based on measurements by
$B$ factories \cite{BaBar:2013mob,BaBar:2012obs,Belle:2016dyj,Belle:2015qfa,Belle:2019rba}
and LHCb \cite{LHCb:2015gmp,LHCb:2017smo,LHCb:2017rln,LHCb:RD2022},
gives $\rD= 0.358 \pm 0.028$ and $\rDs=0.285 \pm 0.013$.
Individually, these values disagree with the SM expectation
\cite{Bigi:2016mdz,Gambino:2019sif,Bordone:2019vic,Bernlochner:2017jka,
Jaiswal:2017rve,Martinelli:2021onb,Bigi:2017jbd,FermilabLattice:2021cdg}
at the significance level of $2.2\sigma$ and $2.3\sigma$, respectively. 
Taken together (the correlation coefficient is -$0.29$),
the disagreement increases to $3.2\sigma$.\footnote{The trend of surplus in tauonic modes is also observed in the measurement of
$R_{J/\psi}$ \cite{LHCb:2017vlu}, the LFU ratio defined similar to \cref{eq:RD_def}
for $B_c\to J/\psi$ transition, while the ratio $R_{\Lambda_c}$
(related to $\Lambda_b\to \Lambda_c$ transition) shows a relative
deficit \cite{LHCb:2022piu}.
These deviations, however, are relatively mild in significance as 
the current associated experimental uncertainties are large.
We  therefore do not include these two ratios in our
analysis. We refer to a recent paper \cite{Fedele:2022iib} analyzing the impact of
$R_{\Lambda_c}$ inclusion in NP analysis.}
On the other hand,  measurements of analogous
muon vs electron LFU ratios are in agreement
with the SM predictions \cite{Belle:2015pkj,Belle:2017rcc,Belle:2018ezy}.

Another set of  anomalies
are observed in $\bsmm$ decays.
One of these is the $\sim 3\sigma$ anomaly
in the measurement of angular observable $P_5^\prime$ in
the decay $\bkstmm$ \cite{LHCb:2013ghj,LHCb:2015svh,LHCb:2020lmf}.
Another sizable tension, of $3.6\sigma$ significance, is 
reported by LHCb~\cite{LHCb:2015wdu,LHCb:2021xxq,LHCb:2021zwz} in the
measurement of branching fraction of the $B_s\to \phi\mu^+\mu^-$ decay,
finding it below the SM expectation.
The data on $B \to K \mu^+\mu^-$ \cite{LHCb:2014cxe} and $\Lambda_b \to \Lambda \mu^+\mu^-$ \cite{LHCb:2015tgy}
also show a deficit in branching fractions with respect to the SM predictions.
The recent update from LHCb \cite{LHCb:2022qnv,LHCb:2022zom} however finds no evidence
of LFU breaking in $\bsll$ decays: the ratios
$\RK ={\mathrm{BR}(B\to K^{(\ast)}\mu^+\mu^-)}/{\mathrm{BR}(B\to K^{(\ast)} e^+ e^-)}$ \cite{Hiller:2003js}
measured in  the dilepton invariant mass bins 
$0.1 <q^2 < 1.1~\mathrm{GeV^2}$ and $1.1 <q^2 < 6~\mathrm{GeV^2}$
are in complete agreement\footnote{It is worth noting that rates of both the 
$B^+\to K^+\mu^+\mu^-$ \cite{LHCb:2014cxe} and $B^+\to K^+ e^+ e^-$ decays in the central $q^2$ bin
 are now low compared to the corresponding SM predictions while their ratio ($R_K$) remains
 SM-like \cite{LHCb:2022zom}.} with the SM predictions known
with percent-level accuracy \cite{Bordone:2016gaq,Isidori:2020acz,Isidori:2022bzw}. These new findings overturn previous
results \cite{LHCb:2017avl,LHCb:2019hip,LHCb:2021trn} that
reported a deficit in $\RK$.
Although the predictions of individual branching fractions and optimized angular observables
such as $P_5^\prime$ are subject to significant hadronic uncertainties
\cite{Matias:2012xw,Das:2012kz,Descotes-Genon:2013vna,Horgan:2013hoa,Beaujean:2013soa,Mandal:2014kma,
Lyon:2014hpa,Jager:2014rwa,Ciuchini:2015qxb,Ciuchini:2021smi,Gubernari:2020eft}, 
at present it is contestable whether the long-distance effects can
fully account for the tension in $\bsll$, and NP may be warranted to explain the data (e.g., see Ref.~\cite{Gubernari:2022hxn}).

Recently, the Fermilab CDF collaboration \cite{CDF:2022hxs}, based on 2002-2011 data
with $8.8~\fbi$ integrated luminosity, reported a new measurement of
$W$-boson mass:
\begin{align}\label{eq:mw_CDF}
	M_W=80.4335 \pm 0.0094 \,\text{GeV},
\end{align}
which differs from the SM prediction
$M_W^\mathrm{SM}=80.357 \pm 0.006$ GeV \cite{Awramik:2003rn} with $7\sigma$ significance.
It is intriguing to note that the CDF measurement also differs with $M_W$ measurements
reported by  ATLAS \cite{ATLAS:2017rzl} and LHCb \cite{LHCb:2021bjt},
an issue to be resolved\footnote{Recently it has been
suggested \cite{Bandyopadhyay:2022bgx} that discrepancies between
different $M_W$ measurements could be due to a light NP particle
modifying the missing transverse momentum in the detector.}
in the future with improved measurements.
Here, we will take a view that the CDF measurement hints towards
NP presence in the $M_W$ value.

In this article we show that a $\Hpm$ boson,
naturally present in simple extensions of the SM
such as two-Higgs doublet model (2HDM) \cite{Lee:1973iz},
can account for the above-mentioned anomalies. 
The fact that $\Hpm$ can explain $\rD$ and $\rDs$ anomalies
is well known in literature \cite{Crivellin:2012ye,Celis:2012dk,Tanaka:2012nw,Ko:2012sv,
Crivellin:2013wna,Crivellin:2015hha,Kim:2015zla,Cline:2015lqp,Lee:2017kbi,
Iguro:2017ysu,Iguro:2018qzf,Chen:2018hqy,Martinez:2018ynq,Li:2018rax,Cardozo:2020uol,
Athron:2021auq,Iguro:2022uzz,Blanke:2022pjy}.
Here, we  show that  same set of $\Hpm$ interactions that
explain $\RD$ anomaly
 unavoidably induce a destructive NP contribution desired to
simultaneously explain the tension in $\bsll$, while
keeping the ratios $\RK$ unaltered.
Our results strengthen the viewpoint that a common NP could
be behind the charged- and neutral-current $B$ anomalies. The 
$M_W$ anomaly can also be explained by a NP contribution
to the Peshkin-Takeuchi $T$ parameter \cite{Peskin:1991sw},
which helps determine the allowed mass range of the physical
scalars in the 2HDM.

\section{\boldmath  $\Hpm$ interactions and relevant NP parameters}
The $\Hpm$ boson  we consider  belongs to a 2HDM without any special
discrete symmetry (see Ref.~\cite{Branco:2011iw} for a comprehensive review).
The  general $\Hpm$ interactions in the fermion mass basis are given by
the Lagrangian \cite{Davidson:2005cw}
\begin{align}\label{eq:Lag}
\cL_{H^+}= -\bar u (V \r^{d} R -  \r^{u\dagger}\, V L )d \, H^+
 -  \bar \nu \, \r^{e} R \,e\, H^+
 + \text{H.c.},
\end{align}
where $\r^{f}$ ($f= u, d, e$) are $3\times 3$ NP Yukawa matrices,
$V$ denotes the CKM matrix, and $L/R\equiv (1\mp \g_5)/2$ are the chirality projectors.
In addition to the $\Hpm$ boson, 2HDM also has CP-even/odd scalar bosons $H$, $A$.
Their Yukawa interactions are not important for our analysis;
we refer to Refs.~\cite{Davidson:2005cw,Mahmoudi:2009zx} for their details.
However we need to define the masses of $H$, $A$,
as those would be required in the computation of $M_W$ in the model.
The masses of $H$, $A$ are related to mass of $\Hpm$ by the relations\footnote{Here we  assume, to
accord with the current data \cite{CMS:2018uag,ATLAS:2019nkf},
that there is very little mixing between SM Higgs ($h$) and $H$ boson.}
\begin{align}\label{eq:m_phi}
	\mHpm^2 = m_H^2 - \frac{v^2}{2}(\Lambda_4+ \Lambda_5),
	\quad m_A^2 = m_H^2 - v^2 \Lambda_5\,,
\end{align}
where $v\simeq 246$ GeV, and $\Lambda_{4}$, $\Lambda_{5}$ are the quartic couplings
in the Higgs potential (see the Appendix for details).

To explain the anomalies with a minimal set of NP parameters,
we make the ansatz that NP Yukawa matrices have the following simple structure
\begin{align}\label{eq:texture}
\r^{u} = 
	\begin{pmatrix}
	0 & 0 & 0\\0 & 0 & 0\\0 & \rtc & 0\\
\end{pmatrix},\quad \r^{e} = 
	\begin{pmatrix}
	0 & 0 & 0\\0 & 0 & 0\\0 & 0 & \rtt\\
\end{pmatrix},
\end{align}
and $\r^{d}=0$. The texture as such in \cref{eq:texture} is the most economical
choice  to  affect rate of
$\bdtn$ only: the off-diagonal coupling $\rtc$ facilitates $\Hpm$ mediated
$b\to c$ transitions that are not CKM suppressed and diagonal lepton coupling $\rtt$ ensures that
only semitauonic modes are affected.
With the above choice the Lagrangian in \cref{eq:Lag} simplifies to
(dropping a $V_{td}$ suppressed term)
\begin{align}\label{eq:Lag2}
	\cL_{H^+} = (\rtc^\ast V_{tb}\,\bar c_Rb_L + \rtc^\ast V_{ts}\bar c_R s_L
			- \rtt \,\bar \nu_{\tau L} \tau_R) H^+
	 		+ \mathrm{H.c.},
\end{align}
which together with  \cref{eq:m_phi}
defines all the Yukawa interactions and NP parameters relevant in our setup.

Concerning direct search constraints on $\Hpm$,
analyses in Refs.~\cite{Iguro:2018fni,Iguro:2022uzz},
based on experimental results of
Refs.~\cite{CMS:2018fza,CMS:2017dcz,ATLAS:2019itm,CMS:2018kcg},
find that mass range  $\mHpm> 400$ GeV for an explanation of the $\RD$ anomaly
is likely ruled out due to a constraint from $pp\to bc \to \tau\nu$ process.
However, the low-mass region $\mHpm < 400$ is still not excluded \cite{Iguro:2022uzz}.
It was pointed out recently \cite{Blanke:2022pjy}
that $\tau\nu$ search with an additional
$b$-tagged jet ($pp\to b H^\pm \to b \tau \nu $) could be useful
in probing this low-mass region of $\Hpm$.
In this article we therefore focus on the $\mHpm < 400$ GeV region.

\section{Observables}
In this section we discuss $\Hpm$ contributions to the anomalous
observables together with  the relevant constraint on our setup. 

\subsection{\boldmath $\rD$ and $\rDs$}
\begin{figure}[t]
\subfloat[]{\includegraphics[width=0.2\textwidth,height=0.1\textheight]{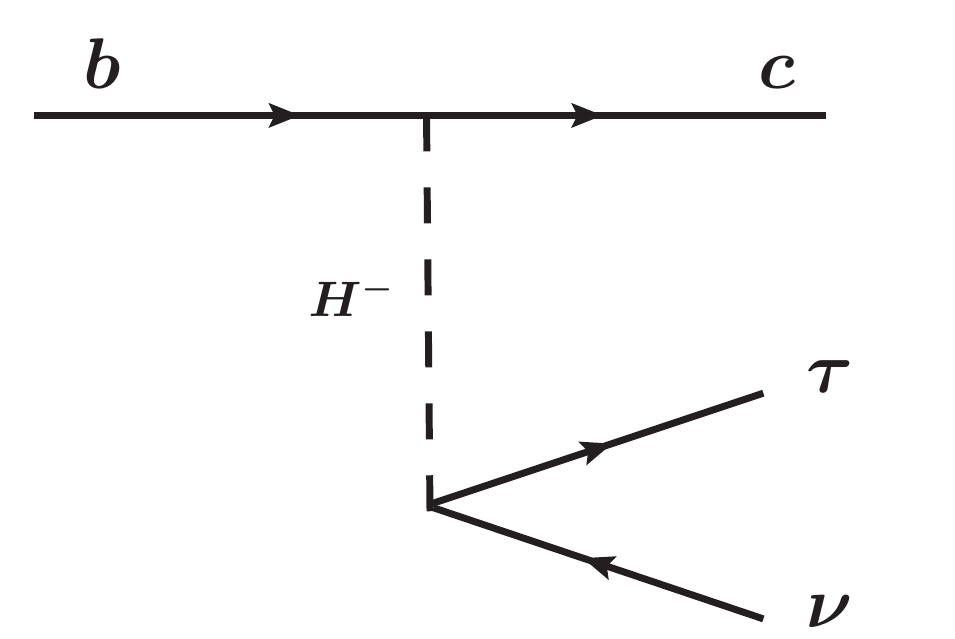}\label{fig:b2ctaunu}}~~~~
\subfloat[]{\includegraphics[width=0.2\textwidth,height=0.1\textheight]{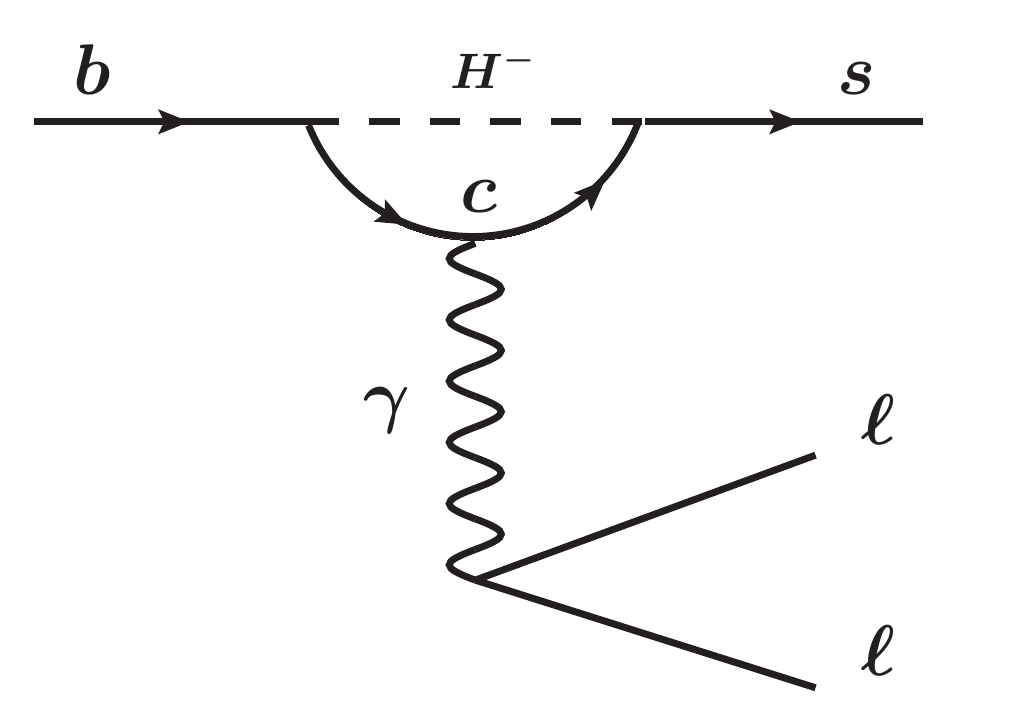}\label{fig:b2sll_penguin}}\\
\subfloat[]{\includegraphics[width=0.2\textwidth, height=0.1\textheight]{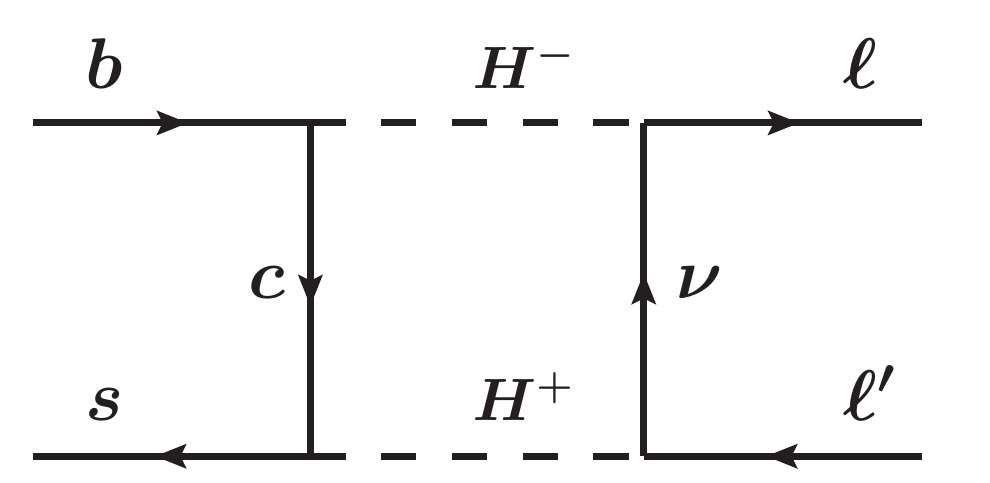}\label{fig:b2sll_box}}~~~~
\subfloat[]{\includegraphics[width=0.2\textwidth,height=0.105\textheight]{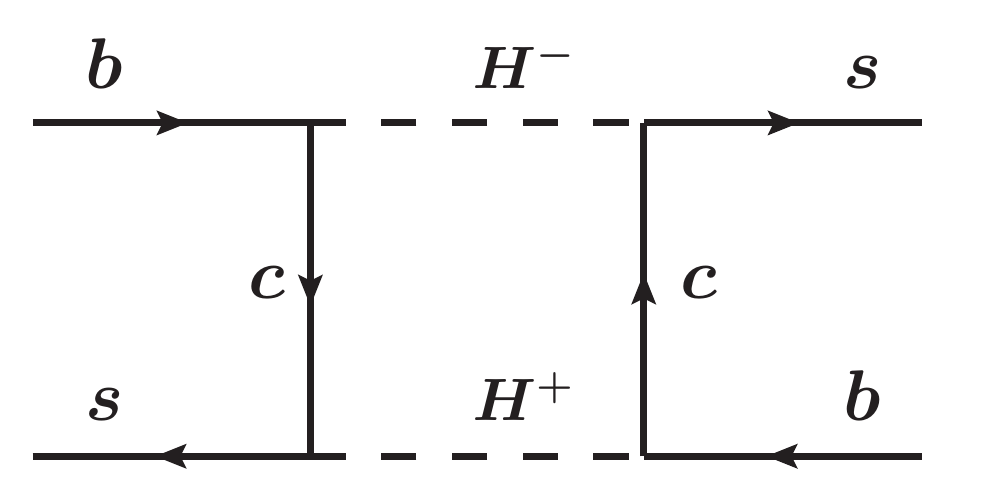}\label{fig:Bmix}}\\
\subfloat[]{\includegraphics[width=0.2\textwidth,height=0.095\textheight]{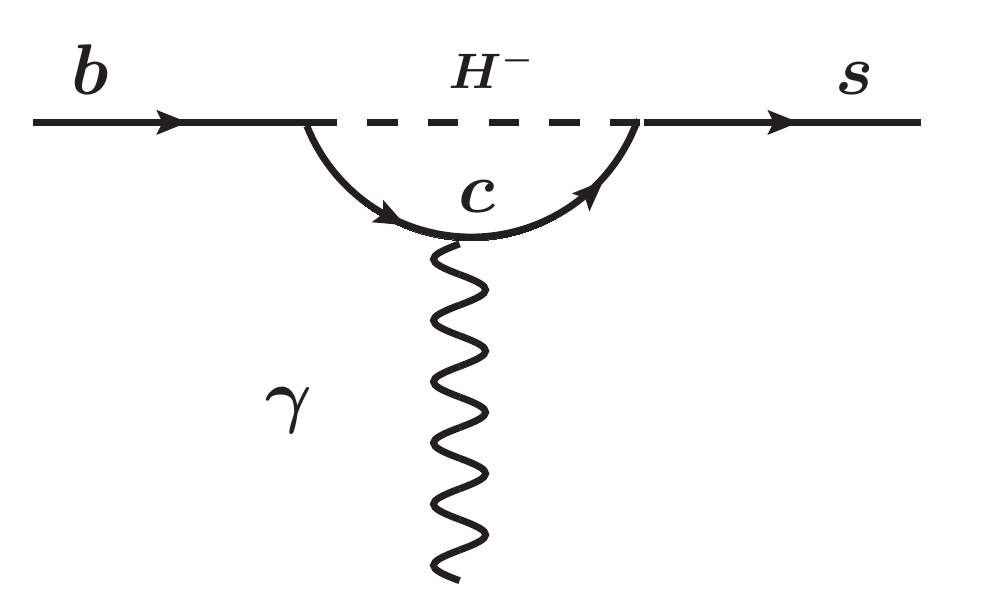}\label{fig:b2sgam}}~~~~
\subfloat[]{\includegraphics[width=0.2\textwidth,height=0.095\textheight]{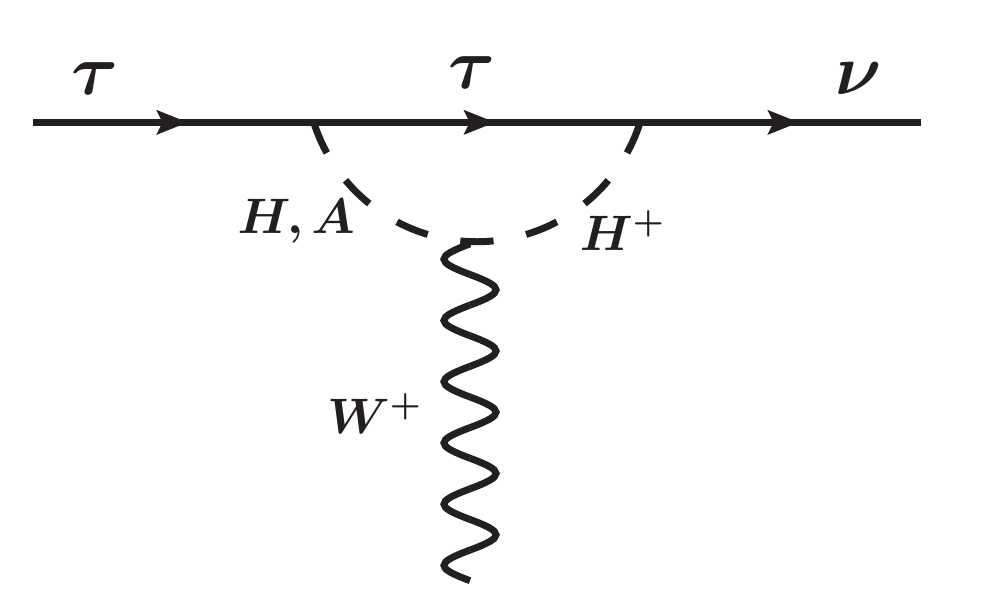}\label{fig:lep_vertex}}
\caption{Feynman diagrams  for $\bctn$ (\ref{fig:b2ctaunu}), $\bsll$ (\ref{fig:b2sll_penguin}, \ref{fig:b2sll_box}), $B_s-\bar B_s$ mixing (\ref{fig:Bmix}), $b \to s \gamma$ (\ref{fig:b2sgam}), and $W$-$\tau$-$\nu$ vertex (\ref{fig:lep_vertex}).}
\label{fig:feyn}
\end{figure}

The $\Hpm$ boson
mediates $\bctn$ transition
at tree-level (shown in \cref{fig:b2ctaunu}), the effect of which can be parametrized by
the following effective Hamiltonian 
\begin{align}\label{eq:effLag_b2ctaunu}
	\cHeff= 2\sqrt{2} G_F V_{cb }\,C_{S, L}(\bar c_R b_L)(\bar \tau_R \nu_{\tau L}),
\end{align}
where the coefficient  $C_{S, L}$ at  scale $\mu \sim \mHpm$ is given by
\begin{align}\label{eq:CSL}
	C_{S, L} = \frac{\rtc^\ast\rtt^\ast }{2\sqrt{2} G_F V_{cb}\,\mHpm^2}.
\end{align}
The contributions of $C_{S, L}$ to ratios $\rD$ and $\rDs$
 are numerically parametrized (at scale $\mu\sim m_b$) as \cite{Iguro:2022yzr}\footnote{Similar formulas
 are also given in Refs.~\cite{Blanke:2018yud,Iguro:2018vqb}.}
\begin{align}\label{eq:RD_num}
	\rD &\simeq (\rD)_\mathrm{SM}[1 + 1.49\operatorname{Re}(C_{S,L})+ 1.01\, |C_{S,L}|^2],\\
	\rDs &\simeq (\rDs)_\mathrm{SM}[1 - 0.11\operatorname{Re}(C_{S,L}) + 0.04|C_{S,L}|^2].
\end{align}

The scalar  interaction in \cref{eq:effLag_b2ctaunu} contributes rather significantly
(due to lack of chirality suppression)
to the $B_c\to \tau \nu$ branching ratio. Numerically, it is given as \cite{Iguro:2022yzr}
\begin{align}
	\text{BR}(B_c \to \tau \nu) \simeq 0.02\, |1 - 4.35\, C_{S,L}|^2.
\end{align}
This decay  is not measured yet.
However, based on the precisely measured lifetime of  
$B_c$ meson \cite{ParticleDataGroup:2022pth}, a
theoretical constraint on maximally allowed $\text{BR}(B_c\to \tau \nu)$
can be obtained \cite{Alonso:2016oyd}.
Recent estimates \cite{Blanke:2018yud,Aebischer:2021ilm} suggest that
$\text{BR}(B_c\to \tau \nu)$ as large as $60\%$ to $63\%$ is still possible.

In our analysis we have not considered  the constraint
from the differential decay distributions of
$\bdtn$ \cite{BaBar:2013mob,Belle:2015qfa}, which are known to be
sensitive to scalar NP
\cite{Sakaki:2014sea,Freytsis:2015qca,Celis:2016azn}.
Compared to ratios $\RD$, the decay distributions 
are quite sensitive to hadronic form factors and parametric (e.g. $V_{cb}$)
uncertainties.
Furthermore, the corresponding experimental analyses \cite{BaBar:2013mob,Belle:2015qfa} are model dependent
and require the NP model's contributions to the background
and the signal efficiency in order to obtain the data.
Also, since the correlations
among different data bins are not available,
a combined data analysis is difficult.
The improved measurements at Belle II \cite{Belle-II:2018jsg}
will be helpful in overcoming these issues (e.g.,
see discussion in Ref.~\cite{Sakaki:2014sea}).
\subsection{\boldmath  $\bsll$}

The $\Hpm$ contributions to $\bsll$ processes have been discussed 
in several works (for example, see 
\cite{Iguro:2017ysu,Iguro:2018qzf,Crivellin:2019dun,Athron:2022afe}),
most of which have focused on top quark-$\Hpm$ loop diagrams.
Such contributions, which are local in the effective field theory at scale $\mu\sim m_b$,
are not present in our setup (see \cref{eq:Lag2}).
Instead, the typical  $\bsll$ contributions  arise from the diagrams involving
charm quark in the loop as shown in \cref{fig:b2sll_penguin,fig:b2sll_box}.

The leading contribution to $\bsll$ comes from the penguin diagram in \cref{fig:b2sll_penguin}.
This contribution in the effective field theory can be obtained via the penguin insertion of the
four-quark operator
$(\bar c_{R}\, b_{L})(\bar s_L \,c_{R})$  mediating $\bscc$ transition.
This four-quark operator is generated at tree level via a diagram similar
to \cref{fig:b2ctaunu} with the $\bar \tau \nu \Hpm$ vertex replaced by $\bar s c \Hpm$.
For convenience we make use of
a Fierz identity
and define the following $\bscc$ effective Hamiltonian
\begin{align}\label{eq:effLag_cc}
	-\cHeff =
	 \frac{4G_F}{\sqrt{2}}V_{tb}V_{ts}^\ast\, \tilde C_{V,LR}\,(\bar s_{L}^\alpha \g^\mu b_{L}^\beta)(\bar c_R^\beta \g_\mu c_{R}^\alpha),
\end{align}
where $\alpha, \beta $ are the color indices,
and the coefficient $\tilde C_{V,LR}$ at scale $\mu \sim \mHpm$ is given as $\tilde C_{V,LR} = - {v^2 }|\rtc|^2/{4 \mHpm^2}$.
Then, closing the charm loop 
of the $\bscc$ operator in \cref{eq:effLag_cc} (diagram shown in \cref{fig:b2sll_penguin} with $\Hpm$ integrated out) effects a nonlocal NP contribution to the vector
operator $(\bar s_L\g_\mu b_L)(\bar \ell \g^\mu \ell)$.
Adapting the results of Refs.~\cite{Jager:2019bgk,Jager:2017gal} to our case, we obtain
the following NP contribution to the  $\bsll$ Wilson coefficient\footnote{We
follow notation of
Ref.~\cite{London:2021lfn} for the $b\to s$ operator basis~\cite{Buchalla:1995vs}.}
\begin{align}\label{eq:c9_cc}
	C_{9 \ell}(q^2, \mu) =
	\left[\frac{4}{9} + h(q^2, m_c, \mu)\right]\tilde C_{V,LR},
	\end{align}
where function $h(q^2, m_c)$ is given in eq.~(11) of Ref.~\cite{Beneke:2001at}.
The \cref{eq:c9_cc} gives sufficiently accurate result\footnote{The dominant contribution comes from $q^2$-independent terms in
$h(q^2, m_c, \mu)$ (see, e.g., Ref.~\cite{Jager:2019bgk}).} if the coefficient
$\tilde C_{V,LR}$ arise at a scale close to the $B$-meson scale.
However, since in our model
the four-quark operator is generated at higher scale $\mu\sim \mHpm$,
the renormalization group (RG) running effects are important.
Therefore, instead of  \cref{eq:c9_cc}, we use \texttt{wilson} package 
\cite{Aebischer:2018bkb} (which
is based on the results of
Refs.~\cite{Jenkins:2013zja,Jenkins:2013wua,Alonso:2013hga,Jenkins:2017jig,
Aebischer:2017gaw,Jenkins:2017dyc,Celis:2017hod,Herren:2017osy}),
accounting for one-loop RG evolution of $\tilde C_{V,LR}(\mu)$,
to evaluate the mixing into $C_{9\ell}(\mu_b)$.
Numerically, taking the NP scale $\mu_\mathrm{high} = 200$ GeV as
an example case,
we find $C_{9\ell}(\mu_b=4.8\,\mathrm{GeV})
= 5.17\,\tilde C_{V,LR}(\mu_\mathrm{high})$.
The $Z$-penguin diagram (\cref{fig:b2sll_penguin} with $\gamma \to Z$)
can be ignored as the corresponding loop function vanishes in the $m_c\to 0$ limit. 

Another contribution to $\bsll$ comes from the box diagram in \cref{fig:b2sll_box} which  gives \cite{Crivellin:2019dun}
\begin{align}\label{eq:c9_box}
	C_{9 \tau}^\text{NP} =C_{10 \tau}^\text{NP}=
	-\frac{v^2}{64 \pi \alpha_e\,\mHpm^2}|\rtc\,\rtt|^2.
\end{align}

The contributions in \cref{eq:c9_cc}
are lepton flavor universal (due to $\g \ell\ell$ vertex).
On the other hand, the contribution in \cref{eq:c9_box} in principle introduces
$\tau$ vs.~$e, \mu$ violation in our setup.
However, this contribution depends on coupling product $|\rtc\,\rtt|$
that, as we will see later, is strongly constrained by the $\bctn$ processes
(and by demanding a solution to the $\RD$ anomaly),
causing contributions in \cref{eq:c9_box} to be completely negligible in
the relevant parameter space.
Consequently, NP contributions to $\bsll$ are practically
described by \cref{eq:c9_cc} and
universal to all lepton flavors.
As a result, in our setup the ratios $\RK$  are SM-like in agreement with the
observations made by the LHCb \cite{LHCb:2022qnv,LHCb:2022zom}.
Equally important to note is that since NP contributions to
Wilson coefficients ($C_{10\ell}$),
related to axial-vector current, are negligible,
the rate of the rare decay $B_s\to \mu^+\mu^-$
remain SM-like, which is also consistent with the
new CMS result \cite{CMS:2022mgd} based on 2016-2018 data corresponding
to integrated luminosity of $140~ \fbi$.
The recent global fit to $\bsll$ data
(excluding $\RK$ and $\mathrm{BR}(B_s\to \mu^+\mu^-)$, which anyway remain
unaffected in the considered scenario)
shows that the NP scenario \cite{Hurth:2022lnw}
\begin{align}\label{eq:c9np}
	C_{9\ell} = -0.95 \pm 0.13
\end{align}
is strongly favored over the no NP hypothesis,
corresponding to  $6.1\,\sigma$ pull away from the SM 
(for other NP scenarios see Ref.~\cite{Greljo:2022jac,Ciuchini:2022wbq}\footnote{For
fits before the new $\RK$ measurements,
see Refs.
\cite{Alok:2019ufo,Datta:2019zca,Hurth:2021nsi,Alguero:2021anc,
Altmannshofer:2021qrr,Ciuchini:2020gvn,Geng:2021nhg,
Alok:2022pjb,SinghChundawat:2022zdf,Biswas:2020uaq}.}).
In our analysis, we will use \cref{eq:c9np} to explain the current $\bsll$ discrepancies.

There are few important flavor constraints on $\rtc$.
The most stringent constraint comes from the
mass difference ($\Delta M_{B_s}$) in $B_s-\bar B_s$ mixing.
The $\Hpm$-induced box diagram
(diagram with $W^+$ and $\Hpm$ in loop vanishes in the $m_c\to 0$ limit),
shown in \cref{fig:Bmix},
gives rise to the effective Hamiltonian,
$\cHeff =  C_{bs}\,(\bar s \gamma^\mu L b)(\bar s \gamma^\mu L b)$, where 
\begin{align}
  C_{bs} = \frac{V_{ts}^{\ast 2}V_{tb}^2 \,|\rtc|^4}
  			{128 \pi^2 m_{H^+}^2}.
\label{eq: C1HH}
\end{align}
The current value of the mass differences is
$\Delta M_{B_s}=17.741\pm 0.020~\text{ps}^{-1}$ \cite{ParticleDataGroup:2022pth},
which is to be compared with the SM prediction
$\Delta M_{B_s}=18.4^{+0.7}_{-1.2}~\text{ps}^{-1}$ \cite{DiLuzio:2019jyq}.

Another relevant constraint arises from  radiative decay $B\to X_s\g$, which gets modified
due to the loop diagram shown in \cref{fig:b2sgam}.
The corresponding dipole coefficient $C_7$ at scale $\mu\sim \mHpm$  at the leading order is given by \cite{Crivellin:2019dun}
 \begin{align}
 	C_{7} = -\frac{7}{36}\frac{v^2}{4 \mHpm^2}|\rtc|^2,
 \end{align}
while the coefficient related to $b \to s g$ is $C_{8} \simeq (6/7) C_7$.
The current experimental value for the  branching ratio of $B\to X_s\gamma$ is 
$ (3.32 \pm 0.15) \times 10^{-4}$ \cite{HFLAV:2019otj}.

There are $\Hpm$ contributions to $K-\bar K$ mixing parameters 
$\varepsilon_K$ and $\Delta M_K$.
The corresponding contributions arise from box diagram shown in \cref{fig:Bmix}
after replacing external quarks $\{bs\} \to \{sd\}$.
We calculate NP contribution to $K-\bar K$ mixing following Ref.~\cite{Hou:2022qvx}
and use  experimental values from Ref.~\cite{ParticleDataGroup:2022pth};
the resulting constraints however turn out to be weaker than those from $B$ physics.

\begin{figure*}[t]
\centering
\includegraphics[scale=0.394]{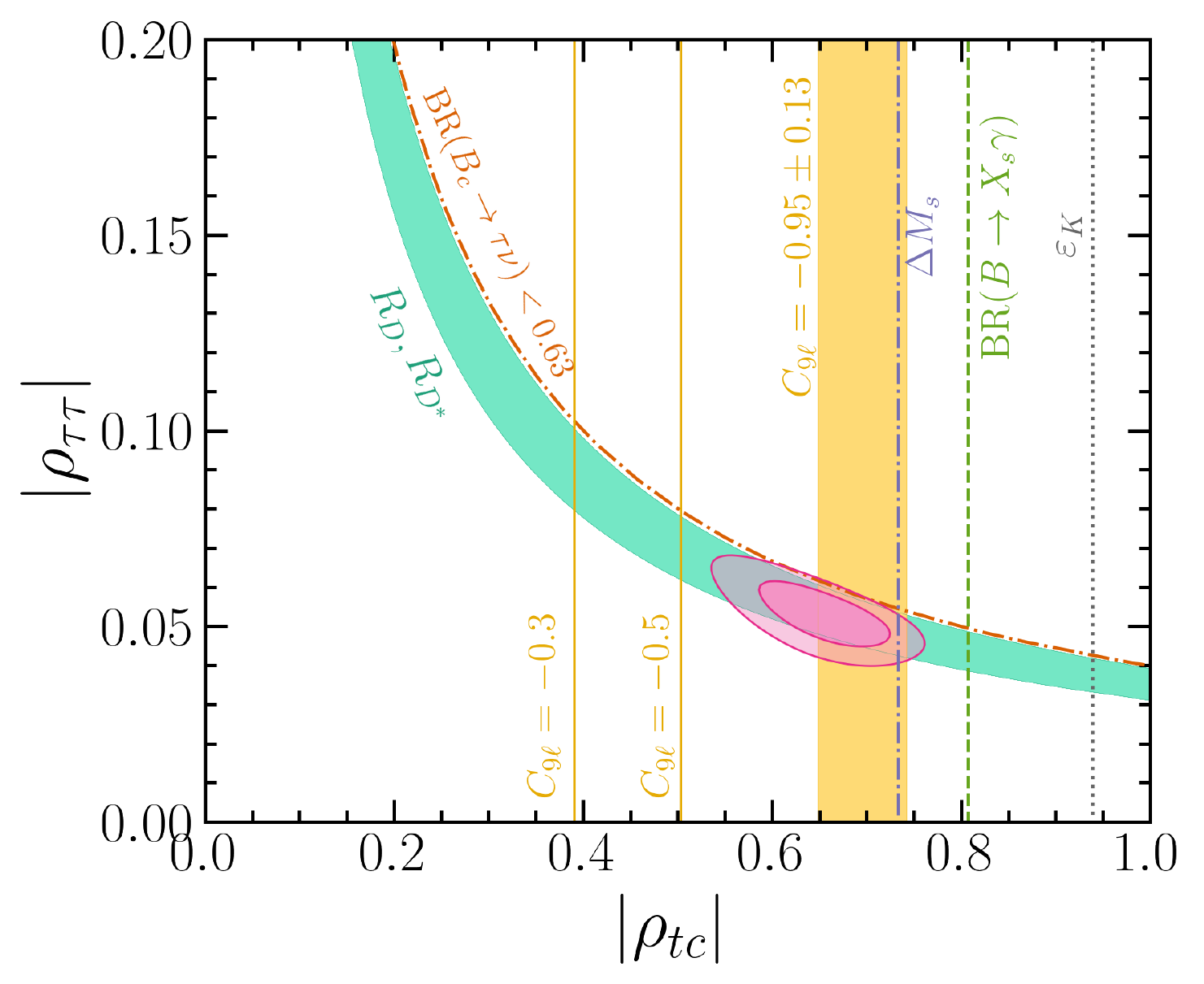}
\includegraphics[scale=0.394]{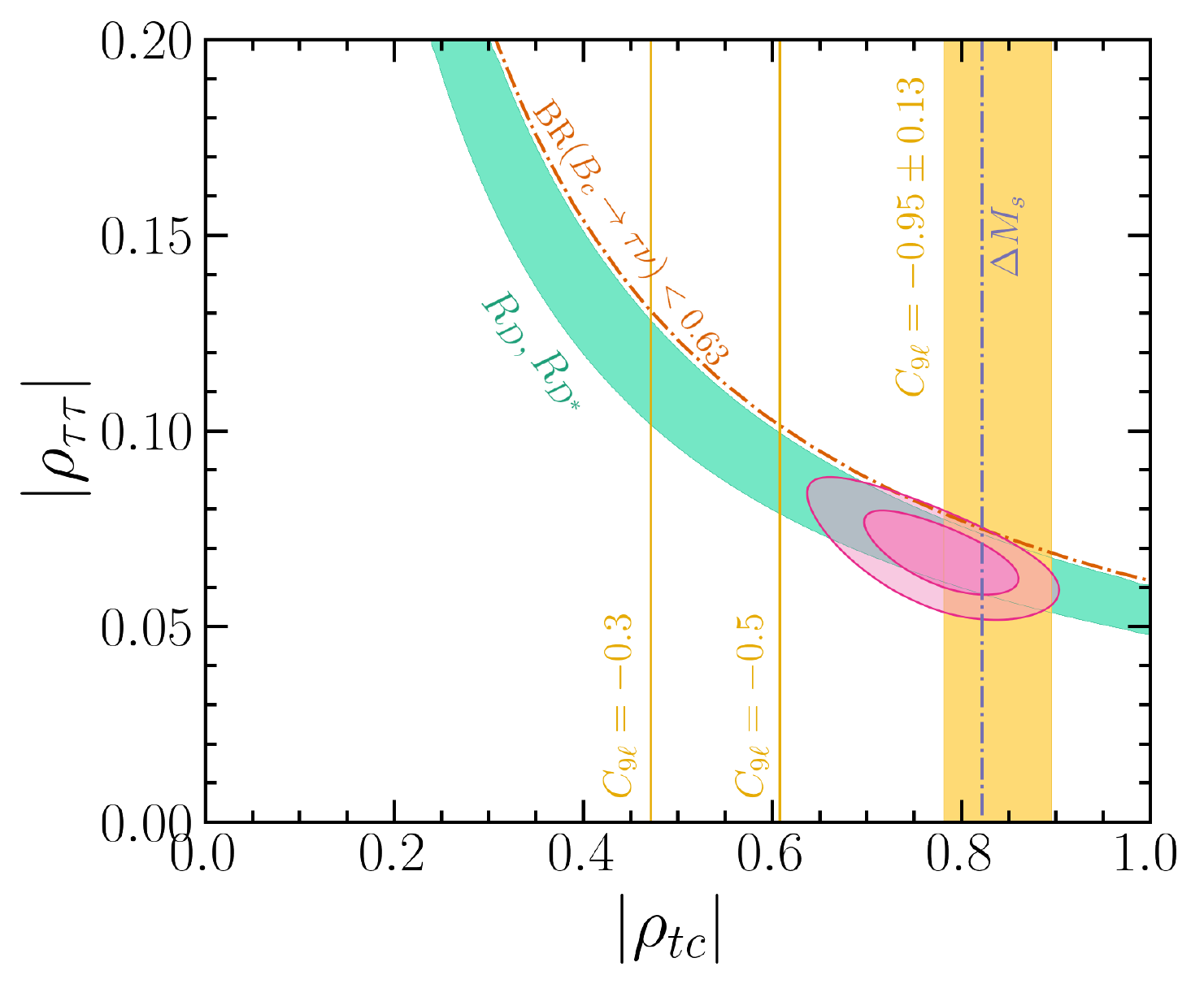}
\includegraphics[scale=0.394]{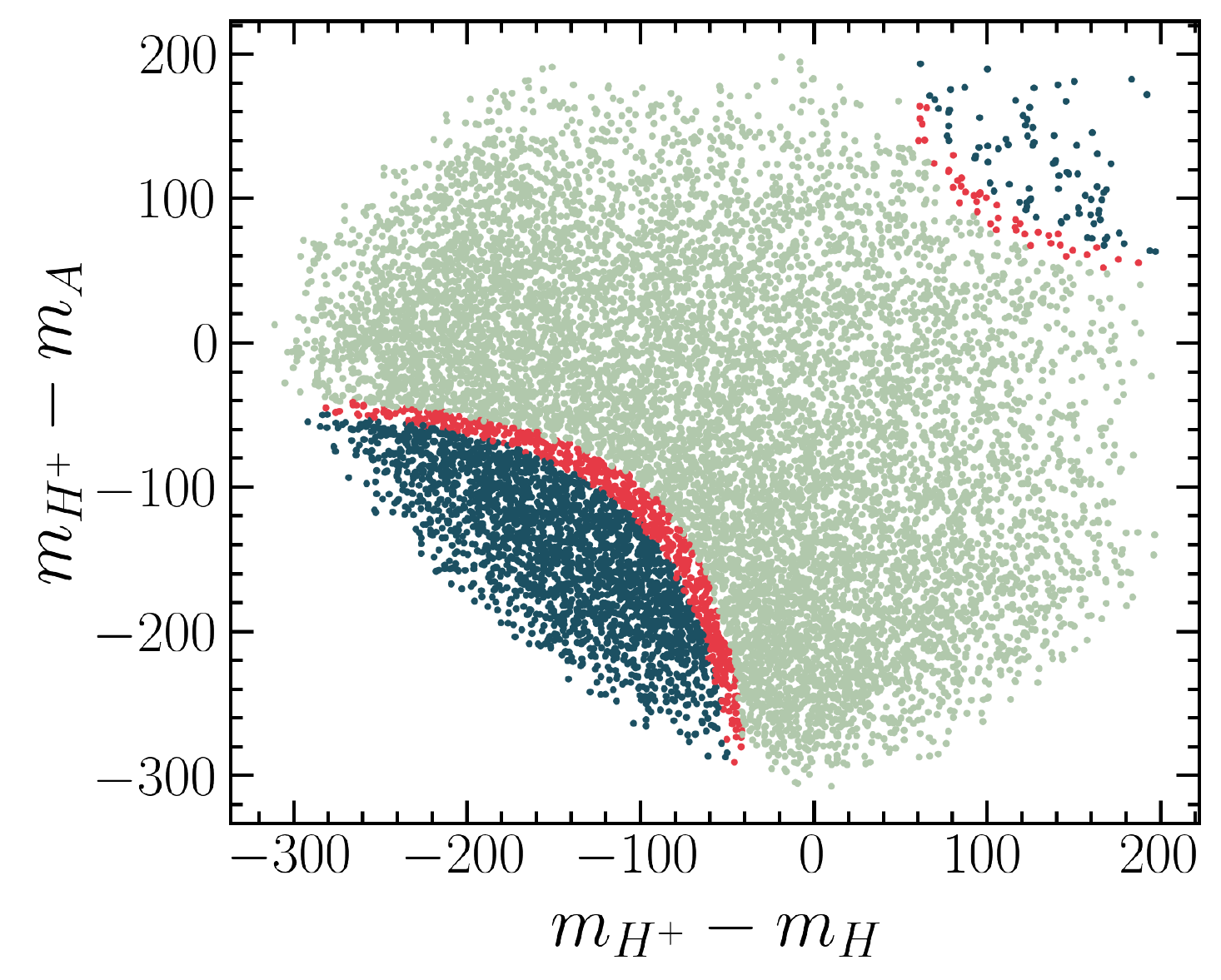}
\caption{First and second plots show results of fits to  $\rD$ and $\rDs$ ($1\sigma$) and
$\bsll$ ($1\sigma$) for $\mHpm=200$ GeV and $250$ GeV, respectively.
The dark  and light magenta color contours are the global $1\sigma $ and $2\sigma$ allowed regions. The third plot shows results of the parameter scan where the red points corresponds to  $M_W$ values that are within $1\sigma$ of \cref{eq:mw_CDF}. See text for other details.}
\label{fig:results}
\end{figure*}

\subsection{\boldmath Shift in $M_W$}
As mentioned in Introduction, the CDF value of $M_W$ differs from the
corresponding SM prediction by $7\sigma$.
This difference can be attributed to a NP correction to
$T$ parameter in the 2HDM (see, e.g.,
Refs.~\cite{Bahl:2022xzi,Song:2022xts,Babu:2022pdn,Arco:2022jrt,Lee:2022gyf,Lu:2022bgw,
Ahn:2022xax,Sakurai:2022hwh,Tran:2022yrh,Arcadi:2022dmt,Ghorbani:2022vtv,Abouabid:2022lpg,
Botella:2022rte,Kim:2022hvh,Hessenberger:2022tcx,Arco:2022jrt}).
The SM value of $M_W$ is calculable as \cite{Babu:2022pdn}
\begin{align}\label{eq:mw_2hdm}
	M_W^2 = \frac{M_Z^2}{2}
	\left(1 + \sqrt{1-\frac{4\pi \alpha_e\,(1+\Delta r)}{\sqrt{2} G_F M_Z^2}}\,\right),
\end{align}
where $\Delta r$ contains quantum corrections associated with oblique parameters
and renormalization of $\alpha_e$. Within the SM, $(\Delta r)_\mathrm{SM}\simeq 0.038$ \cite{Babu:2022pdn}.
Assuming that modifications in $\Delta r$ arise from a NP contribution to $T$ parameter,
one can parametrize NP effects as $\Delta r = (\Delta r)_\mathrm{SM} - (c_W^2/s_W^2)\alpha_e(M_Z)\, T$, where $T$ in 2HDM is given by
\begin{align}\label{eq:T}
	T = \frac{1}{16\pi^2 \alpha_e(M_Z)v^2}\{F(\mHpm^2, m_H^2) + F(\mHpm^2, m_A^2) \nn\\
							- F(m_H^2, m_A^2)\},
\end{align}
with loop function
\begin{align}
	F(a, b) = \frac{a + b}{2} - \frac{ab}{a-b}\log \frac{a}{b}.
\end{align}
Note that $F(a, b)$ vanishes in the limit of $a\to b$, indicating that
at least two of the scalar states should have different  masses in order to
obtain a nonzero contribution to the $T$ parameter.
In our setup, the allowed range of $\mHpm$ is fixed from seeking solution to 
$\RD$ and $\bsll$ anomalies. The values of $m_H$ and $m_A$ then can be obtained
from \cref{eq:m_phi}, with the quartic couplings $\Lambda_{4}$, $\Lambda_{5}$ varied within
perturbative limits. We also include
NP contribution arising from $S$ parameter following Ref.~\cite{Babu:2022pdn};
however these contributions are subdominant.

If $m_H$, $m_A$, and $\mHpm$ are not equal, which is the case to obtain
a finite $T$ parameter as discussed above, then the
vertex $W$-$\tau$-$\nu_\tau$ correction diagram in \cref{fig:lep_vertex}
gives a constraint on $\rtt$ coupling.
This correction is sensitive to the mass splitting of physical scalars in 2HDM and can be parametrized
by writing gauge coupling $g_{W\tau\nu}\to g_{W\tau\nu}(1+\delta g)$,
where $\delta g$ is
\begin{align}
	\delta g  = \frac{|\rtt|^2}{32\pi^2} I(m_H^2/m_{H^\pm}^2, m_A^2/m_{H^\pm}^2),
\end{align}
with loop function $I(x, y)$ given by \cite{Abe:2015oca,Abe:2019bkf},
\begin{equation}
	I(x, y) = 1 + \frac{1}{4}\frac{1+x}{1-x}\log x + \frac{1}{4}\frac{1+y}{1-y}\log y.
\end{equation}
Note that the function $I(x, y)$ vanishes in the combined limit $x\to 1$
and $y\to 1$. The correction $\delta g$ modifies leptonic decay rate of $\tau$ as
$\Gamma_{\tau \to \ell \nu_\tau \bar \nu_{\ell}}
\to\Gamma^\mathrm{SM}_{\tau \to \ell \nu_\tau \bar \nu_{\ell}}(1 + 2 \delta g )$.
The LFU test in $\tau$ decays is then given by ${g_\tau}/{g_e} = 1 + \delta g$,
which is to be compared with the HFLAV  value
$g_\tau/g_e=  1.0029 \pm 0.0014$ \cite{HFLAV:2019otj}.
We note that the $\rtt$ needed in our setup is very small (see next section),
rendering $\delta g$ to be completely negligible  $\sim \mathcal{O}(10^{-5})$ . 
The smallness of $\rtt$ also guarantees that the NP correction (calculated
using the formula given in Ref.~\cite{Cline:2015lqp}) to
the partial leptonic width of $Z\to \tau\tau$ is also negligible.

\section{Results}\label{sec:results}
In our numerical analysis, theoretical predictions of the  flavor observables
are obtained using \texttt{flavio} \cite{Straub:2018kue}.
Our main results  are shown in \cref{fig:results}.
In the first plot, we show results for $\RD$ and $\bsll$
together with relevant constraints in the ($|\rtc|$, $|\rtt|$)
plane for $\mHpm=200$ GeV.
In the plot, the phase $\phi$ ($\equiv \operatorname{arg}(\rtc\rtt)$)\footnote{We take $\rtc$  real so that $\phi$
corresponds to the phase of $\rtt$.}
is fixed by maximizing the global log-likelihood function\footnote{We assume
that experimental uncertainties follow gaussian distribution and for $\mathrm{BR}(B_c\to \tau\bar\nu)$ constraint we assume that it has a uniform probability
within limits $[0, 0.63]$ and zero elsewhere.} in the space of
NP parameters. This  is performed using \texttt{iminuit} \cite{James:1975dr,iminuit}, which
gives the best fit values $|\rtc|=0.659$, $\rtt=0.052$,
$\phi\simeq2\pi/3$.
The green band shows region consistent (within $1\sigma$) with
the current data on $\rD$ and $\rDs$.
The vertical yellow band corresponds to value $C_{9\ell} \sim -1$
($1\sigma$ range of \cref{eq:c9np}, to be exact).
The individual $95\%$ C.L. exclusion bounds
from $\Delta M_{B_s}$, $\text{BR}(B\to X_s\gamma)$, and $\varepsilon_K$
are also shown as vertical lines.
The constraint $\text{BR}(B_c\to \tau\bar\nu)<0.63$ is shown as dash-dotted
orange curve (sitting
just on top of  $1 \sigma$ solution of $\RD$), which rules out the region above it. 
We note that
the $\RD$ solution (green band) only constrains the product $|\rtc\rtt|$,
so the sizes of individual couplings remain unresolved.
Including the data on $\bsll$ decays, which are essentially
sensitive to $|\rtc|$,
a far better constraint on the parameter space is achievable.
The contours in magenta color show $1\sigma$ and $2\sigma$ region where
both $\RD$ and $\bsll$ data can be explained together.
We also show smaller values $C_{9\ell}=-0.5, -0.3$ as solid yellow lines,
illustrating the impact of $\rtc$ variation on NP in $\bsll$.
In the second plot, we show the results for $\mHpm=250$
GeV. The constraints from $B\to X_s\gamma$ and $\varepsilon_K$ are
relaxed and lie outside plot range.
The best fit point now reads $|\rtc|=0.784$, $\rtt=0.068$, and $\phi$ same as before.
In this case we note that $\Delta M_{B_s}$ constraint (dash-dotted blue line)
already covers most of the $1\sigma$
range of \cref{eq:c9np}, but there is still some allowed 
region left. Our results therefore indicate that for $\mHpm>250$ GeV it becomes difficult 
to obtain $C_{9\ell}= -1$,  but
smaller (but  appreciable) values such as $C_{9\ell}\sim -0.5$ are still possible.

In the third plot, we show parameter scan in the plane of
mass-differences $\mHpm-m_H$ and $\mHpm-m_A$, where
the red points corresponds to $M_W$ values  within $1\sigma$ of the 
CDF measurement (\cref{eq:mw_CDF});
the light (dark) green points show $M_W$ values which are below (above)
$1\sigma$ range.
In our setup, as mentioned earlier, the prediction of $M_W$ depends on $\mHpm$
and quartic couplings $\Lambda_4, \Lambda_5$.
To obtain scan results, we vary $\mHpm$ uniformly
in the range ($180$ GeV, $300$ GeV) 
and $\Lambda_4, \Lambda_5$ in the range ($-\sqrt{4\pi}, \sqrt{4\pi}$).
To select allowed points,
we require $m_H^2$, $m_A^2>0$, and reject $m_{H, A}\leq 100$ GeV.
We note that significant population of red points is when both $H, A$
are heavier than $\Hpm$.
There are a few red points in the region when $\Hpm$
is heavier than both $H, A$.
However, we do not find any solution when only one of the $H, A$
is heavier or lighter than $\Hpm$; 
this is because in these corners of parameter space, the NP correction
to $T$ parameter (\cref{eq:T}) is negative, whereas a positive correction
is needed to obtain a positive shift in $M_W$ value. 
In the special case $\Lambda_5=0$, the second relation in \cref{eq:m_phi}
implies $m_H = m_A$, which in \cref{fig:results} (right) corresponds to 
the positive diagonal. So, even though the parameter space is reduced a lot,
$W$-boson mass consistent with the CDF measurement can still be obtained.
On the other hand, in case of vanishing $\Lambda_4$, \cref{eq:m_phi} gives
$\mHpm^2 - m_H^2 = m_A^2 - \mHpm^2$, from which one can deduce that both
$H$, $A$ cannot be simultaneously heavier or lighter than $H^+$,
and therefore from the arguments
presented above we find that the CDF value of $M_W$ will not be explained in
this case.

\section{Conclusions}

At present there are hints of LFU violation
in $b\to c \ell\bar \nu$ data, reinforced further by
the recent LHCb result on the combined measurement of $\rD$ and $\rDs$.
On the other hand, no such effect is seen in $\bsll$, 
and LFU ratios $\RK$ are now SM-like. However, the discrepancies
in the branching fractions and optimized observables related to
$\bsll$ decays still remain.
In this article we show that  a $\Hpm$ boson, of few hundred GeV mass,
can simultaneously explain anomalies in $\RD$ and $\bsll$ data.
That a $\Hpm$ boson can explain the former is already known in literature. 
Here we uncover a nice correlation between $\Hpm$ effects in the
charged- and neutral-current $B$ decays: the enhanced rates of $\bdtn$
imply a destructive NP contribution in the Wilson coefficient $C_{9\ell}$.
We show that the current constraints allow for
$C_{9\ell}\sim -1$, a preferred solution to address discrepancies
in the $\bsll$ decays.
Additionally, we also show that the discrepancy observed in
$M_W$ by CDF II  can also be explained by allowing splitting
among the physical scalars masses;
the solution prefers the neutral states $H, A$ heavier than $\Hpm$.

\vskip0.2cm
\noindent{\bf Acknowledgments} 
I would like to thank Namit Mahajan for a useful conversation and his inputs on the manuscript, and Monika Blanke and Teppei Kitahara for helpful communication regarding their work on charged-current $B$ decays.
I also thank Manu George for reading the paper carefully.
For diagrams and figures in paper, I acknowledge using \texttt{JaxoDraw} \cite{Binosi:2003yf}
and \texttt{Matplotlib} \cite{Hunter:2007}.
This research work is supported by
NSTC 111-2639-M-002-002-ASP of Taiwan. 

\appendix

\begin{widetext}
\section{Scalar potential and mass-spectrum}\label{app:potential}
Here we provide details about the scalar potential of a general 2HDM and
the relations of its parameters with the masses of the scalars in the model.
With $H_1$, $H_2$ denoting Higgs doublets,
the scalar potential is given by \cite{Davidson:2005cw}
\begin{align}\label{eq:potential}
	V(H_1, H_2)=
	M_{11}^{2}|H_1|^{2}
	+M_{22}^{2}|H_2|^{2}
	-\left(M_{12}^{2} H_1^{\dagger} H_2+\mathrm{H.c.}\right)
	 +\frac{\Lambda_{1}}{2}|H_1|^{4}
	 +\frac{\Lambda_{2}}{2}|H_2|^{4}
	 +\Lambda_{3}|H_1|^{2}|H_2|^{2}\nn\\
+~\Lambda_{4}|H_1^{\dagger} H_2|^{2} 
	 +\left\{\frac{\Lambda_{5}}{2}\left(H_1^{\dagger} H_2\right)^{2}\right.
	 +\left[\Lambda_{6}|H_1|^{2}+\Lambda_{7}\left|H_2\right|^{2}\right] H_1^{\dagger} H_2
	 \biggl.+ \,\mathrm{H.c.}\biggr\},
	 \end{align}
where parameters $M_{12}^2$, $\Lambda_{i}$ $(i = 5, 6, 7)$ in general
can have complex phases.
In this paper, we have taken $V(H_1, H_2)$ to be CP-invariant for simplicity,
and therefore all the potential parameters are real.

Working in the Higgs basis \cite{Georgi:1978ri,Lavoura:1994fv,Botella:1994cs}
where only one of the Higgs doublets receives vacuum expectation value,
we define doublets $H_1$, $H_2$
as
\begin{align}
	H_1 = \left(\begin{matrix}
			G^+\\
		\frac{1}{\sqrt{2}}{(v + H_1^0 + i G^0)}\\
	\end{matrix}\right),\quad 
	H_2 = \left(\begin{matrix}
			H^+\\
		\frac{1}{\sqrt{2}}{(H_2^{0} + i A)}\\
	\end{matrix}\right),
\end{align}
so that $\langle H_1\rangle=v/\sqrt{2}$ and $\langle H_2\rangle=0$. 
In the above notation, $G^+$, $G^0$ are the goldstone bosons,
$H^+$ and $A$ are the charged scalar and CP-odd 
scalar, respectively, while the physical CP-even neutral scalars $h$ and $H$
are given by
\begin{align}
	h = H_1^0 \sin \gamma + H_2^{0} \cos\gamma, \quad 
		H = H_1^0 \cos \gamma - H_2^{0} \sin\gamma,
\end{align}
with $\gamma$ denoting the $h$-$H$ mixing angle (analogous to $(\beta - \alpha)$ in
type-II 2HDM notation).

Minimization of the potential gives conditions: $M_{11}^2+\Lambda_1 (v^2/2) =0$, 
$M_{12}^2-\Lambda_6 (v^2/2) =0$. The relations between potential parameters
and the scalar masses, which we are mainly interested in,
are given as \cite{Davidson:2005cw} 
\begin{align}
	m_{H+}^{2} &
	=M_{22}^{2}+\frac{v^{2}}{2} \Lambda_{3}\,,\\
	m_{A}^{2}-m_{H+}^{2} &=
	-\frac{v^{2}}{2}\left(\Lambda_{5}-\Lambda_{4}\right)\,,\\
	 m_{H}^{2}+m_{h}^{2}-m_{A}^{2} &=
	 v^{2}\left(\Lambda_{1}+\Lambda_{5}\right)\,, \\
	 \left(m_{H}^{2}-m_{h}^{2}\right)^{2} &=
	 \left[m_{A}^{2}+\left(\Lambda_{5}-\Lambda_{1}\right) v^{2}\right]^{2}+4 \Lambda_{6}^{2} v^{4}\,,\\
	 \sin \gamma \cos \gamma &=
	 -\frac{\Lambda_{6} v^{2}}{m_{H}^{2}-m_{h}^{2}}.
\end{align}

In case of very small mixing angle, i.e., $\cos\gamma\to 0$, which we have taken
in this paper, the simplified relations for the scalar masses are 
%
	$m_h^2 \simeq \Lambda_1 v^2$, $m_A^2 = M_{22}^2 + (v^2/2)(\Lambda_3 + \Lambda_4 - \Lambda_5)$, 
and the ones given in \cref{eq:m_phi}.
\end{widetext}

\bibliography{bibliography}
\end{document}